\documentclass[a4paper]{jpconf}
\usepackage{graphicx}
\usepackage[hidelinks]{hyperref}
\usepackage{color}
\usepackage{amsmath,amsfonts,amssymb,bm}

\usepackage{lipsum}

\usepackage{fancyhdr}
\def\beq{\begin{equation}}
\def\eeq{\end{equation}}
\def\beqn{\begin{eqnarray}}
\def\eeqn{\end{eqnarray}}

\def\bl {\mbox{\boldmath $[$}}
\def\br {\mbox{\boldmath $]$}}

\def\Q {{\bf Q}}
\def\P {{\bf P}}

\def\bz {\mbox{\boldmath $0$}}
\def\bi {\mbox{\boldmath $i$}}
\def\bo {\mbox{\boldmath $1$}}
\def\bl {\mbox{\boldmath $[$}}
\def\br {\mbox{\boldmath $]$}}
\def\cross {\mbox{\boldmath $\times$}}

\def\x {{\bf x}}
\def\y {{\bf y}}

\def\Q {{\bf Q}}
\def\P {{\bf P}}
\def\R {{\bf R}}

\def\X {{\bf X}}
\def\Y {{\bf Y}}
\def\L {{\bf L}}
\def\lj {{\bf l}}
\def\O {{\bf O}}

\begin{document}

\title{Solvable Model of a Generic Trapped Mixture of
Interacting Bosons: Many-Body and Mean-Field Properties at the Infinite-Particle Limit}

\author{S~Klaiman$^{1}$, A~I~Streltsov$^{1,2}$ and O~E~Alon$^{3}$}
\address{$^{1}$ Theoretical Chemistry, Physical Chemistry Institute, Heidelberg University, Heidelberg, Germany}
\address{$^{2}$ Institute of Physics, University of Kassel, Kassel, Germany}
\address{$^{3}$ Department of Physics, University of Haifa at Oranim, Tivon, Israel}

\ead{ofir@research.haifa.ac.il}

\begin{abstract}
A solvable model of a generic trapped bosonic mixture,
$N_1$ bosons of mass $m_1$ and $N_2$ bosons of mass $m_2$ 
trapped in an harmonic potential of frequency $\omega$ 
and interacting by harmonic inter-particle interactions of strengths
$\lambda_1$, $\lambda_2$, and $\lambda_{12}$,
is discussed.
It has recently been shown for the ground state [J. Phys. A {\bf 50}, 295002 (2017)] that in the infinite-particle limit,
when the interaction parameters $\lambda_1(N_1-1)$, $\lambda_2(N_2-1)$, $\lambda_{12}N_1$, $\lambda_{12}N_2$ are
held fixed,
each of the species is $100\%$ condensed and its density per particle 
as well as the total energy
per particle are given by the solution 
of the coupled Gross-Pitaevskii equations of the mixture.
In the present work we investigate properties of the trapped generic mixture at the infinite-particle limit, 
and find differences between the many-body and mean-field descriptions of the mixture,
despite each species being $100\%$.
We compute analytically 
and analyze, both for the mixture and for each species,
the center-of-mass position and momentum variances, their uncertainty product, the angular-momentum variance, 
as well as the overlap of the exact and Gross-Pitaevskii wavefunctions of the mixture.
The results obtained in this work
can be considered as a step forward in characterizing
how important are many-body effects in a fully condensed trapped 
bosonic mixture at the infinite-particle limit.
\end{abstract}

\section{Introduction}\label{Intro}

Ever since the first experimental demonstration of trapped Bose-Einstein condensates (BECs) in ultra-cold quantum gases
[1-3],
the connection between the microscopic many-particle Hamiltonian and the macroscopic
Gross-Pitaevskii, mean-field theory has drawn much attention
[4-14].
Whereas properties like the energy and density per particle, and being $100\%$ condensed,
are exactly reproduced in the infinite-particle limit by the 
Gross-Pitaevskii theory
[5-8],
other properties, like the variance of many-particle operators and, ultimately, the many-particle wavefunction
itself generally are not reproduced by the mean-field theory [10-14].

Mixtures of BECs have also attracted a lot of interest in their static, thermal, and out-of-equilibrium properties
[15-37],
yet the connection between their 
microscopic many-body Hamiltonian and the mean-field solution for mixtures has only recently received attention
[38-43].
In particular, that each of the species in the ground state of a generic trapped mixture is, in the infinite-particle limit, 
$100\%$ condensed has been recently shown within the exactly-solvable harmonic-interaction model for trapped mixtures \cite{BB_GEN_HIM}.
The harmonic-interaction model has widely been studied for indistinguishable bosons 
[44-52],
fermions
[50-55],
and bosonic mixtures 
[56-60].
The purpose of the present work is to compute and compare
many-body and mean-field properties of the $100\%$ condensed generic mixture \cite{BB_GEN_HIM} at the infinite-particle limit.
The ground-state wavefunction can be prescribed analytically at the exact and mean-field levels,
which facilitates a transparent study of properties of the mixture at both levels of theory.
We extend and broaden previous results obtained in the specific case of a symmetric mixture \cite{BB_HIM}.
We concentrate on the center-of-mass position an momentum variance, their uncertainty product,
and angular-momentum variance of the whole mixture as well as of each species in the mixture.
Generally, in comparison with the textbook single-particle case \cite{QM_book}
the variances and uncertainty product of many-particle operators are more involved
\cite{Variance,TD_Variance,jcs_var}, see in this context also \cite{Drumm,Oriol_Robin}.
Finally, motivated by recent results in the single-species case \cite{SL_Psi,S_INF_Psi},
we evaluate explicitly the overlap of the exact and Gross-Piatesvkii wavefunctions of the generic mixture.
A common line of our investigation is how intra-species and inter-species interaction parameters
(in combination with the masses)
influence the respective properties of the mixture at the infinite-particle limit.

\section{The harmonic-interaction model for a generic trapped mixture}\label{HIM_BB}

We start from the many-particle Hamiltonian ($\hbar=1$)
\beqn\label{HAM_MIX}
& & \hat H(\x_1,\ldots,\x_{N_1},\y_1,\ldots,\y_{N_2}) = \nonumber \\
& & = \sum_{i=1}^{N_1} \left( -\frac{1}{2m_1} \frac{\partial^2}{\partial \x_i^2} + \frac{1}{2} m_1\omega^2 \x_i^2 \right)  
 + \sum_{j=1}^{N_2} \left( -\frac{1}{2m_2} \frac{\partial^2}{\partial \y_j^2} + \frac{1}{2} m_2\omega^2 \y_j^2 \right) + \nonumber \\
& & + 
\lambda_1 \sum_{1 \le i < l}^{N_1} (\x_i-\x_l)^2 + 
\lambda_2 \sum_{1 \le j < m}^{N_2} (\y_j-\y_m)^2 +  
\lambda_{12} \sum_{i=1}^{N_1} \sum_{j=1}^{N_2} (\x_i-\y_j)^2 \
\eeqn
which describes a generic trapped mixture of $N_1$ bosons of mass $m_1$ (species $1$) 
and $N_2$ bosons of mass $m_2$ (species $2$) trapped in an harmonic potential of 
frequency $\omega$ and interacting by harmonic intra-species and inter-species interactions of
strengths $\lambda_1$, $\lambda_2$, and $\lambda_{12}$.
Positive values of $\lambda_1$, $\lambda_2$, and $\lambda_{12}$ mean attraction,
and negative values repulsion.
Of course, a combination of repulsive and attractive intra-species and inter-species interactions is possible \cite{BB_GEN_HIM}.
The Hamiltonian (\ref{HAM_MIX}) can be diagonalized by moving from the laboratory frame 
(i.e., transforming the Cartesian coordinates)
to a set of Jacobi coordinates
\beqn\label{MIX_COOR}
& & \Q_s = \frac{1}{\sqrt{s(s+1)}} \sum_{i=1}^{s} (\x_{s+1}-\x_i), \qquad 1 \le s \le N_1-1,  \nonumber \\
& & \Q_{N_1-1+s} = \frac{1}{\sqrt{s(s+1)}} \sum_{j=1}^{s} (\y_{s+1}-\y_j), \qquad 1 \le s \le N_2-1,  \nonumber \\
& & \Q_{N-1} = \sqrt{\frac{N_2}{N_1}} \sum_{i=1}^{N_1} \x_i - 
                      \sqrt{\frac{N_1}{N_2}} \sum_{j=1}^{N_2} \y_j, \quad
\Q_N = \frac{m_1}{M} \sum_{i=1}^{N_1} \x_i + \frac{m_2}{M} \sum_{j=1}^{N_2} \y_j. \
\eeqn
Note that the Jacobian of the transformation satisfies 
$\left|\frac{\partial(\Q_1,\ldots,\Q_N)}{\partial(\x_1,\ldots,\x_{N_1},\y_1,\ldots,\y_{N_2})}\right|=1$.

The transformed many-particle Hamiltonian is separable and reads
\beqn\label{HAM_DIAG}
& & \hat H(\Q_1,\ldots,\Q_N) = 
\sum_{s=1}^{N_1-1} \left( -\frac{1}{2m_1} \frac{\partial^2}{\partial \Q_s^2} + 
\frac{1}{2} m_1\Omega_1^2 \Q_s^2 \right) + 
\sum_{s=N_1}^{N-2} \left( -\frac{1}{2m_2} \frac{\partial^2}{\partial \Q_s^2} + 
\frac{1}{2} m_2\Omega_2^2 \Q_s^2 \right) + \nonumber \\
& & + \left(-\frac{1}{2M_{12}} \frac{\partial^2}{\partial \Q_{N-1}^2} + 
\frac{1}{2} M_{12} \Omega_{12}^2 \Q_{N-1}^2\right) 
+ \left(-\frac{1}{2M} \frac{\partial^2}{\partial \Q_N^2} + 
\frac{1}{2} M \omega^2 \Q_N^2\right),
\eeqn
where $M_{12} = \frac{m_1m_2}{M}$ and $M=N_1m_1+N_2m_2$ are the reduced and total mass, respectively,
and 
\beqn\label{MIX_FREQ}
& & \Omega_1 = 
\sqrt{\omega^2 + \frac{2}{m_1}\left(\Lambda_1\frac{N_1}{N_1-1}+\Lambda_{21}\right)}, \qquad
 \Omega_2 = \sqrt{\omega^2 + \frac{2}{m_2}\left(\Lambda_2\frac{N_2}{N_2-1}+\Lambda_{12}\right)}, \nonumber \\
& & \Omega_{12} = \sqrt{\omega^2 + 2\left(\frac{\Lambda_{12}}{m_2} + \frac{\Lambda_{21}}{m_1}\right)}, \qquad
 \omega \
\eeqn
are the eigen-frequencies of the decoupled oscillators.
$\Omega_1$ and $\Omega_2$ are associated with the intra-species relative coordinates,
$\Omega_{12}$ with the inter-species relative coordinate 
(between the center-of-mass of species $1$ and the center-of-mass of species $2$),
and the eigen-frequency of the center-of-mass coordinate of the whole mixture is equal to the trapping potential $\omega$.
Herein, the so called mean-field interaction parameters are defined,
$\Lambda_1=\lambda_1(N_1-1)$,
$\Lambda_2=\lambda_1(N_2-1)$,
$\Lambda_{12}=\lambda_{12}N_1$, and $\Lambda_{21}=\lambda_{12}N_2$,
and shall have an important role below.

The ground-state wavefunction of the mixture is thus 
\beqn\label{WAVE_FUN_1}
& & \Psi(\Q_1,\ldots,\Q_N) = 
\left(\frac{m_1\Omega_1}{\pi}\right)^{\frac{3(N_1-1)}{4}}
\left(\frac{m_2\Omega_2}{\pi}\right)^{\frac{3(N_2-1)}{4}}
\left(\frac{M_{12}\Omega_{12}}{\pi}\right)^{\frac{3}{4}}
\left(\frac{M\omega}{\pi}\right)^{\frac{3}{4}} \times \nonumber \\
& & \times
e^{-\frac{1}{2} \left(m_1\Omega_1 \sum_{s=1}^{N_1-1} \Q_s^2 + 
m_2\Omega_2 \sum_{s=N_1}^{N-2} \Q_s^2 +
M_{12}\Omega_{12} \Q_{N-1}^2 + M\omega \Q_N^2 \right)} \
\eeqn
and 
\beqn\label{HIM_MIX_GS_E}
& & E = \frac{3}{2} 
\Bigg[(N_1-1) \sqrt{\omega^2 + \frac{2}{m_1}\left(\Lambda_1\frac{N_1}{N_1-1}+\Lambda_{21}\right)} + \nonumber \\
& & + (N_2-1) \sqrt{\omega^2 + \frac{2}{m_2}\left(\Lambda_2\frac{N_2}{N_2-1}+\Lambda_{12}\right)} 
+ \sqrt{\omega^2 + 2\left(\frac{\Lambda_{12}}{m_2} + \frac{\Lambda_{21}}{m_1}\right)} + \omega \Bigg] \
\eeqn 
is the ground-state energy which can be seen 
as a functional of the numbers and masses of the particles and the interaction parameters.

It is possible to solve the Hamiltonian (\ref{HAM_MIX}) at the mean-field level \cite{BB_GEN_HIM}.
The Gross-Pitaevskii (mean-field) wavefunction is
\beqn\label{MIX_WAV_GP}
& & \Phi^{GP}(\x_1,\ldots,\x_{N_1},\y_1,\ldots,\y_{N_2}) = \nonumber \\
& & = \left(\frac{m_1\Omega_1^{GP}}{\pi}\right)^{\frac{3N_1}{4}} 
\left(\frac{m_2\Omega_2^{GP}}{\pi}\right)^{\frac{3N_2}{4}} 
e^{-\frac{1}{2}\left(m_1\Omega_1^{GP}\sum_{i=1}^{N_1} \x_i^2
+m_2\Omega_2^{GP}\sum_{j=1}^{N_2} \y_j^2\right)}, \nonumber \\
& & 
\Omega_1^{GP}=\sqrt{\omega^2 + \frac{2}{m_1}(\Lambda_1+\Lambda_{21})}, \quad
\Omega_2^{GP}=\sqrt{\omega^2 + \frac{2}{m_2}(\Lambda_2+\Lambda_{12})}. \
\eeqn
It is a double-product wavefunction made of the (interaction-dressed) Gross-Pitaevskii orbitals
\beqn\label{MIX_GP_OR}
& & \phi^{GP}_1(\x) = \left(\frac{m_1\Omega_1^{GP}}{\pi}\right)^{\frac{3}{4}}
e^{-\frac{m_1}{2}\Omega_1^{GP}\x^2}, \quad
\phi^{GP}_2(\y) = \left(\frac{m_2\Omega_2^{GP}}{\pi}\right)^{\frac{3}{4}}
e^{-\frac{m_2}{2}\Omega_2^{GP}\y^2}. 
\eeqn
$\Omega_1^{GP}$ and $\Omega_2^{GP}$ are the Gross-Pitaevskii frequencies.
For $N_1 \gg 1$ and $N_2 \gg 1$ they coincide
with the respective many-body eigen-frequencies $\Omega_1$ and $\Omega_2$ (\ref{MIX_FREQ}).

The mean-field energy per particle is given by
\beqn
& & \!\!\!\!\!\!\!\!\!\!\!\! \varepsilon^{GP} = \frac{E^{GP}}{N} = 
\frac{3}{2(\Lambda_{12}+\Lambda_{21})}
\left[\Lambda_{12}\sqrt{\omega^2 + \frac{2}{m_1}(\Lambda_1+\Lambda_{21})} +
\Lambda_{21}\sqrt{\omega^2 + \frac{2}{m_2}(\Lambda_2+\Lambda_{12})}\right] \
\eeqn
and expressed solely via the masses of the bosons and the interaction parameters for any
number of particles ($N_1$ and $N_2$ and therefore) $N$.
A connection between masses and interaction parameters will further serve below.

\section{Properties at the infinite-particle limit}\label{HIM_BB_Prop}

In the two-species infinite-particle limit,
i.e., when $N_1 \to \infty$ and $N_2 \to \infty$ such that
the interaction parameters 
$\Lambda_1$, $\Lambda_2$, $\Lambda_{12}$, and $\Lambda_{21}$ 
(and therefore $\frac{N_1}{N_2}=\frac{\Lambda_{12}}{\Lambda_{21}}$)
are held fixed (hereafter briefly, the infinite-particle limit) 
we have \cite{BB_GEN_HIM} for the frequencies
$\lim_{N\to \infty} \Omega_1 = \Omega_1^{GP}$ and
$\lim_{N\to \infty} \Omega_2 = \Omega_2^{GP}$,
for the energy per particle
\beq\label{MIX_E_GP_INF}
\lim_{N\to \infty} \frac{E}{N} = 
\varepsilon^{GP},
\eeq
and for the reduced density matrices \cite{Lowdin,Yukalov} per particle
\beqn\label{1B_REDUCED_DENS_LIM}
& & \lim_{N \to \infty} \frac{\rho_1(\x,\x')}{N_1} = \rho_1^{GP}(\x,\x'), \qquad
\lim_{N \to \infty} \frac{\rho_2(\y,\y')}{N_2} = \rho_2^{GP}(\y,\y'),  \
\eeqn
and
\beqn\label{1B_1B_REDUCED_DENS_LIM}
& & \lim_{N \to \infty} \frac{\rho_{12}(\x,\x',\y,\y')}{N_1N_2} = 
\rho_1^{GP}(\x,\x') \rho_2^{GP}(\y,\y'). \
\eeqn
The reduced one-particle density matrices at the Gross-Pitaevskii level are given by
$\rho_1^{GP}(\x,\x') = \phi_1^{GP}(\x) \left\{{\phi_1^{GP}(\x')}\right\}^\ast$ and
$\rho_2^{GP}(\y,\y') = \phi_2^{GP}(\y) \left\{{\phi_2^{GP}(\y')}\right\}^\ast$.
The question we would like to address in this work 
is in what capacity (i.e., for which quantities and by how much) 
the many-body and mean-field solutions are different,
even when each of the species in the 
mixture is $100\%$ condensed [Eq.~(\ref{1B_REDUCED_DENS_LIM})]?

To proceed we need to express the 
Gross-Pitaevskii wavefunction (\ref{MIX_WAV_GP}) 
via the Jacobi coordinates (\ref{MIX_COOR}). 
Using (\ref{Ap1}) and (\ref{Ap2}) we have
\beqn\label{MIX_GP_WF_X_Y_Q}
& & \!\!\!\!\!\!\!\!\! \Phi^{GP}(\Q_1,\ldots,\Q_N) = 
\left(\frac{m_1\Omega_1^{GP}}{\pi}\right)^{\frac{3N_1}{4}}
\left(\frac{m_2\Omega_2^{GP}}{\pi}\right)^{\frac{3N_1}{4}} \times
\nonumber \\
& & \!\!\!\!\!\!\!\!\! \times 
e^{-\frac{m_1}{2}\Omega_1^{GP}
\left[\sum_{s=1}^{N_1-1} \Q_s^2 + \left(\frac{\sqrt{N_2}m_2}{M}\Q_{N-1}+\sqrt{N_1}\Q_N\right)^2 \right]} 
e^{-\frac{m_2}{2}\Omega_2^{GP}
\left[\sum_{s=N_1}^{N-2} \Q_s^2 + \left(-\frac{\sqrt{N_1}m_1}{M}\Q_{N-1} + \sqrt{N_2}\Q_N\right)^2 \right]} = 
\nonumber \\
& & \!\!\!\!\!\!\!\!\! =  
\left(\frac{m_1\Omega_1^{GP}}{\pi}\right)^{\frac{3N_1}{4}}
\left(\frac{m_2\Omega_2^{GP}}{\pi}\right)^{\frac{3N_1}{4}} 
e^{-\frac{m_1}{2}\Omega_1^{GP}\sum_{s=1}^{N_1-1} \Q_s^2} \,
e^{-\frac{m_2}{2}\Omega_2^{GP}\sum_{s=N_1}^{N-2} \Q_s^2} \times
\ \\
& & \!\!\!\!\!\!\!\!\! \times 
e^{-\frac{M_{12}}{2}\left[\frac{m_2N_2}{M}\Omega_1^{GP} +
\frac{m_1N_1}{M}\Omega_2^{GP}\right]\Q_{N-1}^2} 
e^{-\frac{M}{2}\left[\frac{m_1N_1}{M}\Omega_1^{GP} +
\frac{m_2N_2}{M}\Omega_2^{GP}\right]\Q_N^2}
e^{-M_{12}\sqrt{N_1N_2}\left[\Omega_1^{GP}-\Omega_2^{GP}\right]\Q_{N-1}\Q_N}. \nonumber \
\eeqn 
We see that the Gross-Pitaevskii wavefunction is `nearly' separable in terms of the Jacobi coordinates,
except of the last term which couples $\Q_{N-1}$ and $\Q_N$.
Interestingly, along the sector in parameter space in which the mean-field frequencies [see (\ref{MIX_WAV_GP})] are equal,
\beq\label{Sector}
\sqrt{\omega^2 + \frac{2}{m_1}(\Lambda_1+\Lambda_{21})} = 
\sqrt{\omega^2 + \frac{2}{m_2}(\Lambda_2+\Lambda_{12})} \quad \Longleftrightarrow \quad 
m_2(\Lambda_1+\Lambda_{21}) = m_1(\Lambda_2+\Lambda_{12}),
\eeq
the Gross-Pitaevskii wavefunction is separable in terms of the Jacobi coordinates,
as does the exact wavefunction.
Relation (\ref{Sector}) defines
an interaction--mass balance condition between the two species
for the separability of the center-of-mass coordinate
of the mixture at the mean-field level.
Surprisingly, this condition for separability is required even 
in the case the inter-species interaction is zero.
In other words, the center-of-mass coordinate $\Q_N$ of two (harmonic-interacting) trapped BECs 
is not separable at the Gross-Pitaevskii level, 
even if the two species do not interact with each other,
unless $m_2\Lambda_1 = m_1\Lambda_2$.
We note that condition (\ref{Sector}) obviously holds for the symmetric mixture \cite{BB_HIM}. 
On the other hand, the Gross-Pitaevskii wavefunction is separable in terms of the
center-of-mass coordinates of the 
individual species (see below),
whereas the exact wavefunction is not.

We now express the position and momentum center-of-mass operators
of the mixture and of each of the species in terms of the Jacobi coordinates.
Thus we have
\beq\label{RCM}
\hat \R_{CM} = \Q_N, \quad \hat \P_{CM} = \frac{1}{i}\frac{\partial}{\partial \Q_N}, \quad
\bl\hat \R_{CM},\hat \P_{CM}\br = \bi \quad \forall N
\eeq
for the mixture
($\bi$ and similarly $\bo$ below are shorthand symbols for $i$ and $1$ 
in each of the three Cartesian components)
and
\beqn\label{XYCM}
& & \hat \X_{CM} = \sqrt{\frac{N_2}{N_1}}\frac{m_2}{M}\Q_{N-1}+\Q_N, \quad
\hat \P_{X_{CM}} = \frac{1}{i}\frac{\partial}{\partial \X_{CM}} =
\sqrt{N_1N_2} \frac{1}{i}\frac{\partial}{\partial \Q_{N-1}} + 
\frac{N_1m_1}{M}\frac{1}{i}\frac{\partial}{\partial \Q_N}, \nonumber \\
& & \bl\hat \X_{CM},\hat \P_{\X_{CM}}\br = \bi, \quad \forall N, \nonumber \\ 
& & \!\!\!\!\!\!\!\!\!\!\!\! \hat \Y_{CM} = -\sqrt{\frac{N_1}{N_2}}\frac{m_1}{M}\Q_{N-1} + \Q_N, \quad
\hat \P_{Y_{CM}} = \frac{1}{i}\frac{\partial}{\partial \Y_{CM}} =
-\sqrt{N_1N_2}\frac{1}{i}\frac{\partial}{\partial \Q_{N-1}} + \frac{N_2m_2}{M}\frac{1}{i}\frac{\partial}{\partial \Q_N}, \nonumber \\
& & \!\!\!\!\!\!\!\!\!\!\!\! \bl\hat \Y_{CM},\hat \P_{\Y_{CM}}\br = \bi, \quad \forall N, 
\eeqn
for each of the species.
The center-of-mass operators of the species and mixture are related by the relations
\beq
 \frac{m_1N_1 \hat \X_{CM} + m_2N_2 \hat \Y_{CM}}{M} = \hat \R_{CM}, \quad
\hat \P_{X_{CM}} + \hat \P_{Y_{CM}} = \hat \P_{CM}
\eeq
which are to be used 
in the computation of the variances at the mean-field level.

In what follows we use a compact notation to denote the variance (of the three Cartesian components) 
of an operator $\hat \O = (\hat O_1, \hat O_2, \hat O_3)$
with respect to the exact and Gross-Pitaevskii wavefunctions,   
\beqn
& & \!\!\!\!\!\!\!\!\!\!\!\!
\Delta^2_{\hat \O} = \langle\Psi|\hat \O^2|\Psi\rangle - \langle\Psi|\hat \O|\Psi\rangle^2, \qquad
\Delta^2_{\hat \O,GP} = \langle\Phi^{GP}|\hat \O^2|\Phi^{GP}\rangle - \langle\Phi^{GP}|\hat \O|\Phi^{GP}\rangle^2,
\eeqn 
where $\hat \O^2 \equiv (\hat O^2_1, \hat O^2_2, \hat O^2_3)$.

The variances of the center-of-mass position and momentum 
operators of the mixture, at the exact many-body level,
are given by
\beqn\label{VAR_MIX}
& & \Delta^2_{\hat \R_{CM}} = 
\frac{1}{M} \frac{1}{2\omega} \bo, \qquad
\Delta^2_{\hat \P_{CM}} = 
M \frac{\omega}{2} \bo, \quad \forall N. \
\eeqn
The position variance decreases with the mass of the mixture, 
whereas the momentum variance increases with the mass.
Otherwise, they do not depend on other parameters of the mixture, like the interaction parameters for instance.
Consequently, their uncertainty product
\beq\label{UP_MIX}
\Delta^2_{\hat \R_{CM}} \Delta^2_{\hat \P_{CM}} = \frac{1}{4} \bo, \quad \forall N \
\eeq
holds for any number of particles.
The uncertainty product is minimal,
reflecting the separability of the center-of-mass 
of the mixture in the harmonic trap.

On the other hand, at the Gross-Pitaevskii level we find
\beqn
& & \!\!\!\!\!\!\!\!\!\!\!\! \Delta^2_{\hat \R_{CM},GP} = 
\frac{1}{M}\left[\frac{m_1N_1}{M}\frac{1}{\sqrt{1 + \frac{2}{m_1\omega^2}(\Lambda_1+\Lambda_{21})}} + 
\frac{m_2N_2}{M}\frac{1}{\sqrt{1 + \frac{2}{m_2\omega^2}(\Lambda_2+\Lambda_{12})}}\right] 
\frac{1}{2\omega} \bo,  \\
& & \!\!\!\!\!\!\!\!\!\!\!\! \Delta^2_{\hat \P_{\R_{CM}},GP} = 
M\left[\frac{m_1N_1}{M}\sqrt{1 + \frac{2}{m_1\omega^2}(\Lambda_1+\Lambda_{21})} + 
\frac{m_2N_2}{M}\sqrt{1 + \frac{2}{m_2\omega^2}(\Lambda_2+\Lambda_{12})}\right] 
 \frac{\omega}{2} \bo, \quad \forall N. \nonumber \
\eeqn
The center-of-mass momentum (position) variance again scales with the (inverse) mass of the system,
but depends explicitly on the products of the relative mass of each species times
the corresponding (inverse) 
Gross-Pitaevskii frequency.
The variances depend explicitly on all parameters of the mixture, in particular on the interaction parameters. 
This is entirely different than the many-body dependence (\ref{VAR_MIX}), 
and reflects the dressing of the frequency (and inseparability) 
of the center-of-mass coordinate at the mean-field level.
Consequently, 
the uncertainty product 
\beqn
& &
\Delta^2_{\hat \R_{CM},GP} \Delta^2_{\hat \P_{CM},GP} = 
 \left\{1 + \frac{m_1m_2\Lambda_{12}\Lambda_{21}}{(m_1\Lambda_{12}+m_2\Lambda_{21})^2} 
\left[\frac{\sqrt{1 + \frac{2}{m_1\omega^2}(\Lambda_1+\Lambda_{21})}}
{\sqrt{1 + \frac{2}{m_2\omega^2}(\Lambda_2+\Lambda_{12})}} + \right.\right. \nonumber \\
& & + \left.\left.
\frac{\sqrt{1 + \frac{2}{m_2\omega^2}(\Lambda_2+\Lambda_{12})}}
{\sqrt{1 + \frac{2}{m_1\omega^2}(\Lambda_1+\Lambda_{21})}} -2 \right]\right\} 
 \frac{1}{4} \bo \ge \frac{1}{4} \bo, \quad \forall N \
\eeqn
is larger than the minimal,
except when the interaction--mass balance condition (\ref{Sector}) holds and
the Gross-Pitaevskii wavefunction (\ref{MIX_GP_WF_X_Y_Q}) 
becomes separable with respect to the center-of-mass coordinate.
For instance, increasing either of the intra-species interaction parameters $\Lambda_1$ or $\Lambda_2$
leads to an increase of the uncertainty product.
The above described discrepancy between the mean-field and exact results can be used
as a characterization and measure of the many-body contribution to the physics of a given
$100\%$-condensed trapped mixture. 

We now move to the properties of the center-of-mass operators of the species.
Using relations (\ref{XYCM}) in terms of the Jacobi coordinates we find
\beqn
& & \Delta^2_{\hat \X_{CM}} = 
\frac{1}{m_1N_1} \frac{1}{2\omega} 
\left[\frac{m_1N_1}{M} + \frac{m_2N_2}{M}\frac{1}{\sqrt{1+\frac{2}{\omega^2}\left(\frac{\Lambda_{12}}{m_2}+\frac{\Lambda_{21}}{m_1}\right)}}\right]\bo, \nonumber \\
& & \Delta^2_{\hat \P_{\X_{CM}}} =
m_1N_1 \frac{\omega}{2} 
\left[\frac{m_1N_1}{M} + \frac{m_2N_2}{M}\sqrt{1+\frac{2}{\omega^2}\left(\frac{\Lambda_{12}}{m_2}+\frac{\Lambda_{21}}{m_1}\right)}\right]\bo, \quad \forall N, \nonumber \\
& & \Delta^2_{\hat \Y_{CM}} = 
\frac{1}{m_2N_2} \frac{1}{2\omega} 
\left[\frac{m_2N_2}{M} + \frac{m_1N_1}{M}\frac{1}{\sqrt{1+\frac{2}{\omega^2}\left(\frac{\Lambda_{12}}{m_2}+\frac{\Lambda_{21}}{m_1}\right)}}\right]\bo, \nonumber \\
& & \Delta^2_{\hat \P_{\Y_{CM}}} =
m_2N_2 \frac{\omega}{2} 
\left[\frac{m_2N_2}{M} + \frac{m_1N_1}{M}\sqrt{1+\frac{2}{\omega^2}\left(\frac{\Lambda_{12}}{m_2}+\frac{\Lambda_{21}}{m_1}\right)}\right]\bo, \quad \forall N. \
\eeqn
The center-of-mass position and momentum variances of each species depend only on the
inter-species 
interaction parameters $\Lambda_{12}$ and $\Lambda_{21}$, 
and not on the
intra-species
parameters $\Lambda_1$ and $\Lambda_2$.
Increasing the inter-species interaction, such that the species attract each other more, 
decreases (increases) the position (momentum) variance, and vice versa.
Generally, the variances of species $1$ are different than those of species $2$
because of the mass--particle imbalance (i.e., $m_1N_1 \ne m_2N_2$) of the species. 
Interestingly, however, 
the two uncertainty products are equal
\beqn\label{XYCM_UN}
& & \Delta^2_{\hat \X_{CM}} \Delta^2_{\hat \P_{\X_{CM}}} = 
\Delta^2_{\hat \Y_{CM}} \Delta^2_{\hat \P_{\Y_{CM}}} = 
\left\{1 + \frac{m_1m_2\Lambda_{12}\Lambda_{21}}{(m_1\Lambda_{12}+m_2\Lambda_{21})^2} 
\left[\frac{1}{\sqrt{1+\frac{2}{\omega^2}\left(\frac{\Lambda_{12}}{m_2}+\frac{\Lambda_{21}}{m_1}\right)}}
 + \right.\right. \nonumber \\
& & + \left.\left.
\sqrt{1+\frac{2}{\omega^2}\left(\frac{\Lambda_{12}}{m_2}+\frac{\Lambda_{21}}{m_1}\right)} -2 \right]\right\} 
 \frac{1}{4} \bo > \frac{1}{4} \bo, \quad \forall N. \
\eeqn
The uncertainty products of the individual species are always larger than (the minimal uncertainty product) $\frac{1}{4}$,
unless the inter-species interaction is zero, and we have two decoupled systems.
The dependence on the mass--particle imbalance is then absorbed.
In the specific case of a symmetric mixture, 
these expressions boil down to those given in \cite{BB_HIM}. 

At the mean-field level, the Gross-Pitaevskii wavefunction (\ref{MIX_WAV_GP},\ref{MIX_GP_WF_X_Y_Q})
is separable in terms of
the center-of-mass coordinates of each species and we thus have
\beqn
& & \Delta^2_{\hat \X_{CM},GP} = \frac{1}{m_1N_1} \frac{1}{2\omega} 
\frac{1}{\sqrt{1 + \frac{2}{m_1\omega^2}(\Lambda_1+\Lambda_{21})}}
\bo, \nonumber \\
& & \Delta^2_{\hat \P_{\X_{CM}},GP} = 
m_1N_1 \frac{\omega}{2} \sqrt{1 + \frac{2}{m_1\omega^2}(\Lambda_1+\Lambda_{21})} \bo, \quad \forall N, \nonumber\\
& & \Delta^2_{\hat \Y_{CM},GP} = \frac{1}{m_2N_2} \frac{1}{2\omega} 
\frac{1}{\sqrt{1 + \frac{2}{m_2\omega^2}(\Lambda_2+\Lambda_{12})}} \bo, \nonumber \\
& & \Delta^2_{\hat \P_{\Y_{CM}},GP} = 
m_2N_2 \frac{\omega}{2} \sqrt{1 + \frac{2}{m_2\omega^2}(\Lambda_2+\Lambda_{12})} \bo, \quad \forall N. \
\eeqn
Again, the center-of-mass position and momentum variances of species $1$ are generally different
than those of species $2$.
However, at the mean-field level 
they depend on both the intra-species
and inter-species interaction parameters,
rather than only on the inter-species interaction parameters.
Furthermore, the (non-trivial) dependence on the mass--particle imbalance is absent.
The two uncertainty products are again equal, 
\beqn\label{XYCM_UN_GP}
& & \Delta^2_{\hat \X_{CM},GP} \Delta^2_{\hat \P_{\X_{CM}},GP} = 
\Delta^2_{\hat \Y_{CM},GP} \Delta^2_{\hat \P_{\Y_{CM}},GP} = \frac{1}{4} \bo, \quad \forall N, \
\eeqn
but, in contrast to the exact relation (\ref{XYCM_UN}), are actually minimal.
This is because of the (artificial) separation of the center-of-mass coordinates
of the individual species in the Gross-Pitaevskii wavefunction.
This result adds to our above conclusion, i.e., that in addition to the uncertainty product of the whole mixture
also the uncertainty products of the individual species can be used for characterizing and defining the many-body contributions
to the physics of a $100\%$-condensed trapped mixture. 
 
We now move to discuss the variance of the angular momentum in the mixture at both
the exact and mean-field levels.
To identify the contribution to the angular momentum at the level of a single particle, individual species, and the whole mixture,
\beq\label{AML1}
 \hat \lj_{1,i'} = \frac{1}{i} \x_{i'} \cross \frac{\partial}{\partial \x_{i'}}, \quad
 \hat \lj_{2,j} = \frac{1}{i} \y_j \cross \frac{\partial}{\partial \y_j}, \quad
 \hat \L_1 = \sum_{i=1}^{N_1} \hat \lj_{1,i}, \quad 
 \hat \L_2 = \sum_{j=1}^{N_2} \hat \lj_{2,j}, \quad 
 \hat \L = \hat \L_1 + \hat \L_2,
\eeq
it is instrumental to exploit the representation of the wavefunctions in the laboratory frame (Cartesian coordinates).
The mean-field wavefunction (\ref{MIX_WAV_GP}) is a `spherically symmetric' double-product state.
It hence satisfies at the level of a single particle 
$\hat \lj_{1,i} \Phi^{GP}=\bz \Phi^{GP}$ and $\hat \lj_{2,j} \Phi^{GP}=\bz \Phi^{GP}$.
Consequently,
\beqn
& & \Delta^2_{\hat \L_1,GP} = \Delta^2_{\hat \L_2,GP} = \Delta^2_{\hat \L,GP} = \bz, \quad \forall N.
\eeqn
For the exact wavefunction we recall its structure in the laboratory frame \cite{BB_GEN_HIM}
\beqn\label{WAVE_FUN_XY}
& & \!\!\!\!\!\!
\Psi(\x_1,\ldots,\x_{N_1},\y_1,\ldots,\y_{N_2}) = \left(\frac{m_1\Omega_1}{\pi}\right)^{\frac{3(N_1-1)}{4}}
\left(\frac{m_2\Omega_2}{\pi}\right)^{\frac{3(N_2-1)}{4}}
\left(\frac{M_{12}\Omega_{12}}{\pi}\right)^{\frac{3}{4}}
\left(\frac{M\omega}{\pi}\right)^{\frac{3}{4}} \times \nonumber \\
& & \!\!\!\!\!\!
\times 
e^{-\frac{\alpha_1}{2} \sum_{j=1}^{N_1} \x_j^2 - \beta_1 \sum_{1 \le j < k}^{N_1} \x_j \cdot \x_k} \,
e^{-\frac{\alpha_2}{2} \sum_{j=1}^{N_2} \y_j^2 - \beta_2 \sum_{1 \le j < k}^{N_2} \y_j \cdot \y_k} \,
e^{+\gamma \sum_{j=1}^{N_1} \sum_{k=1}^{N_2} \x_j \cdot \y_k},  \
\eeqn
where the inter-species coupling constant is 
\beq
\gamma = M_{12}(\Omega_{12}-\omega) = 
\frac{\omega}{\frac{N_1}{m_2}+\frac{N_2}{m_1}}\left(\sqrt{1 + \frac{2}{\omega^2}
\left(\frac{\Lambda_{12}}{m_2} + \frac{\Lambda_{21}}{m_1}\right)}-1\right).
\eeq
The other constants $\alpha_1$, $\beta_1$, $\alpha_2$, and $\beta_2$ appearing in (\ref{WAVE_FUN_XY})
are not needed for the computation of the angular momentum and given in \cite{BB_GEN_HIM}.
Examining the structure of (\ref{WAVE_FUN_XY}) we see that
the $\alpha_1$, $\alpha_2$ terms alone are eigenfunctions of $\hat \lj_{1,i}$ and $\hat \lj_{2,j}$, respectively,
the $\beta_1$, $\beta_2$ terms alone are eigenfunctions of $\hat \L_1$ and $\hat \L_2$, respectively,
and the $\gamma$ term is an eigenfunction of $\hat \L$.
All in all, whereas $\hat \L \Psi=\bz \Psi$, 
only $\langle\Psi|\hat \lj_{1,i}|\Psi\rangle=0$, $\langle\Psi|\hat \lj_{2,j}|\Psi\rangle=0$
and 
$\langle\Psi|\hat \L_1|\Psi\rangle=0$ and $\langle\Psi|\hat \L_2|\Psi\rangle=0$
hold for the exact ground state.
This is unlike the structure and above properties 
of the Gross-Pitaevskii wavefunction (\ref{MIX_WAV_GP}).
Of course,
\beqn
& & \Delta^2_{\hat \L} = \bz, \quad \forall N.
\eeqn
Since $\Psi$ is not an eigenfunction of either $\hat \L_1$ or $\hat \L_2$ (unless the inter-species interaction is zero and the species decoupled),
fluctuations are expected.
Making use of the structure of (\ref{WAVE_FUN_XY}) we have
\beqn\label{L1_1}
& & \hat \L_1 \Psi = - \hat \L_2 \Psi = \frac{1}{i} \gamma N_1N_2 (\hat \X_{CM} \cross \hat \Y_{CM}) \Psi = \nonumber \\
& & = \frac{1}{i} \gamma N_1N_2 \left[\left(\sqrt{\frac{N_2}{N_1}}\frac{m_2}{M}\Q_{N-1}+\Q_N\right) \cross 
\left(-\sqrt{\frac{N_1}{N_2}}\frac{m_1}{M}\Q_{N-1} + \Q_N\right)\right] \Psi = \nonumber \\
& & = \frac{1}{i} \gamma \sqrt{N_1N_2} (\Q_{N-1} \cross \Q_N) \Psi.
\eeqn
Thus, for the variance of the intra-species angular momentum operators $\hat \L_1$ and $\hat \L_2$ we find
\beqn\label{L1_2}
& & \Delta^2_{\hat \L_1} =  \Delta^2_{\hat \L_2} = 
\gamma^2 N_1N_2 \langle\Psi| (\Q_{N-1} \cross \Q_N)^2 |\Psi\rangle = 
\nonumber \\
& & = \frac{1}{2} 
\frac{\frac{\Lambda_{12}}{m_2}\frac{\Lambda_{21}}{m_1}}
{\left(\frac{\Lambda_{12}}{m_2}+\frac{\Lambda_{21}}{m_1}\right)^2}
\frac{\left[\sqrt{1 + \frac{2}{\omega^2}\left(\frac{\Lambda_{12}}{m_2} + \frac{\Lambda_{21}}{m_1}\right)}-1\right]^2}{\sqrt{1+\frac{2}{\omega^2}\left(\frac{\Lambda_{12}}{m_2}+\frac{\Lambda_{21}}{m_1}\right)}}\bo, \quad \forall N, \
\eeqn
where we have used the separability of $\Psi$ in terms of the Jacobi coordinates and 
the spherical symmetry of the problem.
The variance is independent of the number of particles and is always non-zero for non-zero inter-species interaction.
On the other hand,
if we examine the variances 
of the intra-species angular-momentum operators
per particle,
$\frac{\hat \L_1}{N_1}$ and $\frac{\hat \L_2}{N_2}$,
we find from $\Delta^2_{\frac{\hat \L_1}{N_1}} = \frac{1}{N_1^2}\Delta^2_{\hat \L_1}$,
$\Delta^2_{\frac{\hat \L_2}{N_2}} = \frac{1}{N_2^2}\Delta^2_{\hat \L_2}$
that they vanish in the infinite-particle limit.
In other words,
at the single-particle level
the angular-momentum each boson carries in the infinite-particle limit vanishes,
just like the mean-field behavior.
It would be instructive to study then the variance of the appropriate conjugate 
angular-momentum (angle) variable \cite{RMP_CONJ}
in the mixture,
which is left for further investigations.

The exact and Gross-Pitaevskii wavefunctions are given analytically above for any number 
$N_1$, $N_2$ 
of particles in the mixture,
and are obviously different from each other even in the limit $N$ goes to infinity.
Motivated by recent results in the single-species case \cite{SL_Psi,S_INF_Psi},
we would like to study the overlap between the Gross-Pitaevskii and exact wavefunctions
which,
by virtue of their difference, 
must be smaller than $1$.

For the final result of the overlap we find
\beqn\label{overlap1}
& & \!\!\!\!\!\!\!\!\!
S_{12}(N_1,N_2,\Lambda_1,\Lambda_2,\Lambda_{12},\Lambda_{21}) = \langle\Phi^{GP}|\Psi\rangle = \nonumber \\
& & \!\!\!\!\!\!\!\!\!
= 2^{\frac{3}{2}N} 
\frac{\Omega_1^{\frac{3}{4}(N_1-1)}[\Omega_1^{GP}]^{\frac{3}{4}(N_1-1)}}
{\left[\Omega_1+\Omega_1^{GP}\right]^{\frac{3}{2}(N_1-1)}}
\frac{\Omega_2^{\frac{3}{4}(N_2-1)}[\Omega_2^{GP}]^{\frac{3}{4}(N_2-1)}}
{\left[\Omega_2+\Omega_2^{GP}\right]^{\frac{3}{2}(N_2-1)}} \times \\
& &  \!\!\!\!\!\!\!\!\! 
\times 
\frac{[\Omega_1^{GP}]^{\frac{3}{4}}[\Omega_2^{GP}]^{\frac{3}{4}}\Omega_{12}^{\frac{3}{4}}\omega^{\frac{3}{4}}}
{\left[\!\left(\Omega_{12}\!+\!\frac{m_2N_2}{M}\Omega_1^{GP}\!+\!\frac{m_1N_1}{M}\Omega_2^{GP}\right)\!\!
\left(\omega\!+\!\frac{m_1N_1}{M}\Omega_1^{GP}\!+\!\frac{m_2N_2}{M}\Omega_2^{GP}\right)\!-\!\frac{m_1N_1m_2N_2}{(m_1N_1+m_2N_2)^2}\!\left(\Omega_1^{GP}\!-\Omega_2^{GP}\right)^2\right]^{\frac{3}{2}}}. \nonumber 
\
\eeqn
The overlap depends on all interaction parameters in the mixture and explicitly on the number of particles in each species. 
We can now perform the infinite-particle limit. The final result reads
\beqn\label{overlap2}
& & \!\!\!\!\!\!\!\!\!
\lim_{N \to \infty} S_{12}(N_1,N_2,\Lambda_1,\Lambda_2,\Lambda_{12},\Lambda_{21}) = 
\nonumber \\
& & \!\!\!\!\!\!\!\!\!
= 2^3
\frac{\left[{1 + \frac{2}{m_1\omega^2}(\Lambda_1+\Lambda_{21})}\right]^{\frac{3}{8}}
\left[{1 + \frac{2}{m_2\omega^2}(\Lambda_2+\Lambda_{12})}\right]^{\frac{3}{8}}
\left[1 + \frac{2}{\omega^2}\left(\frac{\Lambda_{12}}{m_2} + \frac{\Lambda_{21}}{m_1}\right)\right]^{\frac{3}{8}}}
{\left[AB-C^2\right]^{\frac{3}{2}}} \
\eeqn
with
\beqn\label{overlap3}
& & \!\!\!\!\!\!\!\!\!\!\!\!\!\! A = 
\left[1 + \frac{2}{\omega^2}\left(\frac{\Lambda_{12}}{m_2} + \frac{\Lambda_{21}}{m_1}\right)\right] +
\frac{m_2\Lambda_{21}}{m_1\Lambda_{12}+m_2\Lambda_{21}} \left[{1 + \frac{2}{m_1\omega^2}(\Lambda_1+\Lambda_{21})}\right] + \nonumber \\
& & \!\!\!\!\!\!\!\!\!\!\!\!\!\! + \frac{m_1\Lambda_{12}}{m_1\Lambda_{12}+m_2\Lambda_{21}} \left[{1 + \frac{2}{m_2\omega^2}(\Lambda_2+\Lambda_{12})}\right],
\nonumber \\
& & \!\!\!\!\!\!\!\!\!\!\!\!\!\! B = 1 +
\frac{m_1\Lambda_{12}}{m_1\Lambda_{12}+m_2\Lambda_{21}} \left[{1 + \frac{2}{m_1\omega^2}(\Lambda_1+\Lambda_{21})}\right] + \frac{m_2\Lambda_{21}}{m_1\Lambda_{12}+m_2\Lambda_{21}} \left[{1 + \frac{2}{m_2\omega^2}(\Lambda_2+\Lambda_{12})}\right],
\nonumber \\
& & \!\!\!\!\!\!\!\!\!\!\!\!\!\! C = \frac{\sqrt{m_1m_2\Lambda_{12}\Lambda_{21}}}{m_1\Lambda_{12}+m_2\Lambda_{21}}\left\{\left[{1 + \frac{2}{m_1\omega^2}(\Lambda_1+\Lambda_{21})}\right] - \left[{1 + \frac{2}{m_2\omega^2}(\Lambda_2+\Lambda_{12})}\right]\right\}.
\
\eeqn
At the infinite-particle limit, 
the overlap depends on the interaction parameters only and, as expected, 
is always smaller than $1$ in presence of either intra-species or inter-species interactions.
At this limit, the overlap boils down to a product of two single-species overlaps 
(at the infinite-particle limit) provided there is no inter-species interaction
{\it and} then only when the interaction--mass balance condition (\ref{Sector}) holds and the mean-field frequencies are equal.
This separability relation between the overlap in the mixture and the overlaps
within each of its species, in the infinite-particle limit, reads
$S_{12}(\Lambda_1,\frac{m_2}{m_1}\Lambda_1,0,0)=S_1(\Lambda_1) S_2(\frac{m_2}{m_1}\Lambda_1)$,
with obvious notation for the quantities.
This concludes our study.

\section{Summary}\label{Sum}

We have considered in the present work a generic trapped mixture of Bose-Einstein condensates
whose ground-state wavefunction can be prescribed analytically at the exact and mean-field levels.
This situation greatly facilitates a transparent and comparative 
study of properties of the mixture at both levels of theory.
We have derived in this work general expressions for any number of particle in the mixture,
and have concentrated on the infinite-particle limit
in which both wavefunctions admit the same energy per particle 
and $100\%$ condensation of each of the species.

We have computed and 
investigated the center-of-mass position and momentum variances, their uncertainty product,
and the angular-momentum variance of each of the species in the mixture as well as of the whole mixture.
Particular attention has been paid at identifying how differences between 
the exact and mean-field wavefunctions
lead to explicit deviations between the respective quantities.
The renormalization of the center-of-mass frequency and that of
the relative coordinate between the center-of-mass of each species,
as well as the (artificial) coupling between these two Jacobi coordinates
and the (artificial) decoupling between the center-of-mass of each species, all in the Gross-Pitaevskii wavefunction, 
lead to discrepancies between the exact and mean-field properties.
At the bottom line, the two different wavefunctions must have a lower than $1$ overlap,
which has been computed explicitly and analytically as well. 
The results obtained in this study
can be considered as a step forward in characterizing
how important are many-body effects in a generic trapped 
$100\%$-condensed bosonic mixture.

As a brief outlook, 
it would be interesting to study more involved uncertainty relations \cite{pra3,prl3} in a generic mixture,
to extend the investigations on the relation of the exact and mean-field solutions to excited states, when possible,
and to explore the further opportunities hiding in trapped multi-species mixtures.

\ack

This research was supported by the Israel Science Foundation (Grant No.~600/15). 
We thank Kaspar Sakmann and Lorenz Cederbaum for discussions.
OEA is grateful to the continuous hospitality of the Lewiner Institute for Theoretical
Physics (LITP) at the Department of Physics, Technion.

\appendix\section{Transformation between the quadratic terms 
in Cartesian coordinates (laboratory frame) and Jacobi coordinates}\label{APP_JAC}

The position terms are related by
\beqn\label{Ap1}
& & \sum_{i=1}^{N_1} \x_i^2 = \sum_{s=1}^{N_1-1} \Q_s^2 
+ \left(\frac{\sqrt{N_2}m_2}{M}\Q_{N-1}+\sqrt{N_1}\Q_N\right)^2 = \nonumber \\
& & = \sum_{s=1}^{N_1-1} \Q_s^2 
+ \frac{N_2m_2^2}{M^2} \Q_{N-1}^2 + N_1 \Q_N^2 
+ \frac{2\sqrt{N_1N_2}m_2}{M} \Q_{N-1} \Q_N, 
\nonumber \\
& & \sum_{j=1}^{N_2} \y_j^2 = \sum_{s=N_1}^{N-2} \Q_s^2 
+ \left(-\frac{\sqrt{N_1}m_1}{M}\Q_{N-1} + \sqrt{N_2}\Q_N\right)^2 = \nonumber \\
& & = \sum_{s=N_1}^{N-2} \Q_s^2 
+ \frac{N_1m_1^2}{M^2}\Q_{N-1}^2 + N_2 \Q_N^2 
- \frac{2\sqrt{N_1N_2}m_1}{M} \Q_{N-1} \Q_N \
\eeqn
and the momentum terms by
\beqn\label{Ap2}
& & \sum_{i=1}^{N_1} \frac{\partial^2}{\partial \x_i^2} = 
\sum_{s=1}^{N_1-1} \frac{\partial^2}{\partial \Q_s^2} + 
\left(\sqrt{N_2} \frac{\partial}{\partial \Q_{N-1}} + 
\frac{\sqrt{N_1}m_1}{M}\frac{\partial}{\partial \Q_N}\right)^2 = \nonumber \\
& & = 
\sum_{s=1}^{N_1-1} \frac{\partial^2}{\partial \Q_s^2} + 
N_2 \frac{\partial^2}{\partial \Q_{N-1}^2} + 
\frac{N_1m_1^2}{M^2}\frac{\partial^2}{\partial \Q_N^2} + 
\frac{2\sqrt{N_1N_2}m_1}{M} \frac{\partial}{\partial \Q_{N-1}} \frac{\partial}{\partial \Q_N}, \nonumber \\
& & \sum_{j=1}^{N_2} \frac{\partial^2}{\partial \y_j^2} = 
\sum_{s=N_1}^{N-2} \frac{\partial^2}{\partial \Q_s^2} + 
\left(-\sqrt{N_1}\frac{\partial}{\partial \Q_{N-1}} + 
\frac{\sqrt{N_2}m_2}{M} \frac{\partial}{\partial \Q_N}\right)^2 = \nonumber \\
& & = 
\sum_{s=N_1}^{N-2} \frac{\partial^2}{\partial \Q_s^2} + 
N_1 \frac{\partial^2}{\partial \Q_{N-1}^2} + 
\frac{N_2m_2^2}{M^2} \frac{\partial^2}{\partial \Q_N^2} - 
\frac{2\sqrt{N_1N_2}m_2}{M} \frac{\partial}{\partial \Q_{N-1}} \frac{\partial}{\partial \Q_N}. \
\eeqn


\end{document}